\documentclass{iaus}
\usepackage{graphicx}
\usepackage{upmath}

\title[Asteroseismic stellar parameters] 
{Fundamental stellar properties from asteroseismology}

\author[V. Silva Aguirre, L. Casagrande, \& A. Miglio]   
{V\'ictor Silva Aguirre$^1$, Luca Casagrande$^2$ \and Andrea Miglio$^3$}

\affiliation{$^1$ Stellar Astrophysics Centre, Department of Physics and Astronomy, Aarhus University, Ny Munkegade 120, DK-8000 Aarhus C, Denmark \\ email: {\tt victor@phys.au.dk} \\[\affilskip]
$^2$ Research School of Astronomy and Astrophysics, Mount Stromlo Observatory, The Australian National University, ACT 2611, Australia \\ email: {\tt luca@mso.anu.edu.au} \\
[\affilskip]
$^3$ School of Physics and Astronomy, University of Birmingham, Birmingham, B15 2TT, UK \\email: {\tt  miglioa@bison.ph.bham.ac.uk }}

\pubyear{2014}
\volume{298}  
\pagerange{XX -- YY}
\jname{IAUS\,298 Setting the scence for Gaia and LAMOST}
\editors{S. Feltzing, G. Zhao, N.\,A. Walton \& P.\,A. Whitelock, eds.}
\begin{document}

\maketitle
\begin{abstract}
Accurate characterization of stellar populations is of prime importance to correctly understand the formation and evolution process of our Galaxy. The field of asteroseismology has been particularly successful in such an endeavor providing fundamental parameters for large samples of stars in different evolutionary phases. We present our results on determinations of masses, radii, and distances of stars in the CoRoT and Kepler fields, showing that we can map and date different regions of the galactic disk and distinguish gradients in the distribution of stellar properties at different heights. We further review how asteroseismic determinations can produce a unique set of constraints, including ages, outside the solar neighborhood for galactic chemical evolution models. \keywords{stars: fundamental parameters, stars: oscillations, Galaxy: structure}
\end{abstract}
\firstsection 
\section{Introduction}
Accurate and precise parameters of stars are a fundamental cornerstone in astrophysics. To understand the intricate details of physical process occurring in stellar interiors, as well as study galactic archeology from composite populations, knowledge of basic properties of stars is required. Thanks to the revolution led by the CoRoT (\cite[Baglin et al. (2006)]{ab06}) and \textit{Kepler} (\cite[Gilliland et al.(2010)]{gill10}) missions in the field of asteroseismology, precise masses, radii, and ages can now be determined for field stars oscillating via the same mechanism as the Sun (solar-like oscillators, see \cite[Chaplin \& Miglio (2013)]{cm13}).
 
Global stellar properties can be determined from asteroseismology using the two main features of the power spectra of a given solar-like oscillator. The frequencies of oscillation are approximately equally spaced by a characteristic separation, called the large frequency separation $\Delta\nu$, which is a proxy of the mean density of the star (\cite[Ulrich (1986)]{ru86}). The overall pulsation spectra is contained within a gaussian-shaped envelope, whose frequency of maximum power $\nu_{\mathrm{max}}$ scales with the star's gravity and effective temperature (e.g., \cite[Brown et al. (1991)]{tb91}). In this manner, two asteroseismic relations scaled from the solar values can be written to determine mass and radius (e.g., \cite[Kallinger et al.(2010)]{tk10}):
\begin{equation}\label{eqn:mass} 
\frac{M}{M_\odot} \simeq \left(\frac{\nu_{\mathrm{max}}}{\nu_{\mathrm{max},\odot}}\right)^{3} \left(\frac{\Delta\nu}{\Delta\nu_\odot}\right)^{-4}\left(\frac{T_\mathrm{eff}}{T_{\mathrm{eff},\odot}}\right)^{3/2}, 
\end{equation}
\begin{equation}\label{eqn:rad} 
\frac{R}{R_\odot} \simeq \left(\frac{\nu_{\mathrm{max}}}{\nu_{\mathrm{max},\odot}}\right) \left(\frac{\Delta\nu}{\Delta\nu_\odot}\right)^{-2}\left(\frac{T_\mathrm{eff}}{T_{\mathrm{eff},\odot}}\right)^{1/2}. 
\end{equation}
Here, $T_{\mathrm{eff},\odot}= 5777\,\rm K$, $\Delta\nu_\odot = 135.1\pm0.1\,\rm \mu Hz$ and $\nu_{\mathrm{max},\odot}= 3090\pm30\,\rm \mu Hz$ are the observed values in the Sun (\cite[Huber et al. (2011)]{dh11}). These relations can be applied to the thousands of star with oscillations detected by the space missions (e.g., \cite[Miglio et al. (2009)]{am09}, \cite[Chaplin et al. (2011)]{wc11}).

One critical requirement for employing the above relations is an accurate determination of the effective temperature. To accomplish this we make use of the InfraRed Flux method (IRFM), originally developed to determined angular diameters with high accuracy (\cite[Blackwell \& Shallis (1977)]{bs77}). Briefly, the IRFM aims at simultaneously recovering the bolometric and infrared monochromatic flux of a star measured at the top of earth's atmosphere, and compare them to the same theoretical quantities. This reduces to a proper derivation of stellar fluxes, which in turns results in a determination of $T_\mathrm{eff}$ and therefore also in the angular diameter.

As the IRFM requires a value of stellar gravity to determine $T_\mathrm{eff}$, we have coupled the \cite[Casagrande et al. (2006), (2010)]{lc06,lc10} implementation of the IRFM to the asteroseismic scaling relations as described in \cite[Silva Aguirre et al. (2011a)]{vsa11a}. When a determination of metallicity is also available, use of stellar tracks can further constrain the stellar parameters and give an estimation of the stellar age (e.g., \cite[Gai et al. (2011)]{ng11}). We apply this method in the manner described in \cite[Silva Aguirre et al. (2012), (2013)]{vsa12,vsa13}, obtaining as a byproduct of this technique a determination of self-consistent stellar distances from the relation between radius and angular diameter.
\section{Results}
\begin{figure}[th]
\begin{center}
 \includegraphics[width=\textwidth]{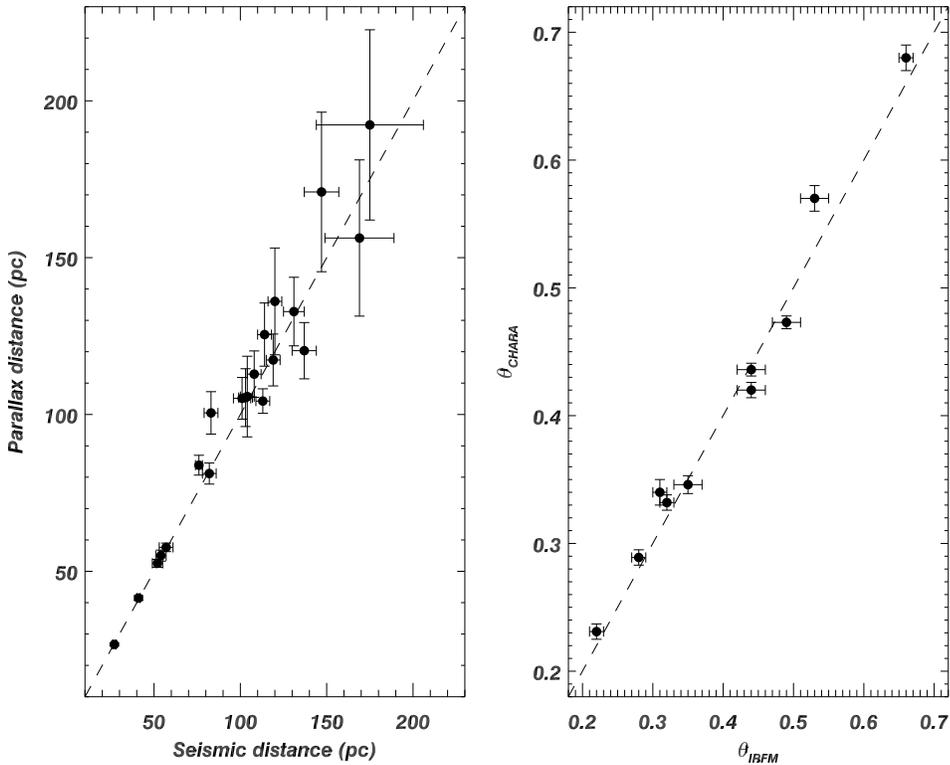} 
 \caption{\textit{Left panel:} comparison of the distances determined with asteroseismology and the IRFM to those from parallax measurements. Sample from \cite[Silva Aguirre et al. (2012)]{vsa12}. \textit{Right panel:} comparison of angular diameters from asteroseismology and the IRFM to those from interferometric measurements made by \cite[Huber et al.(2012)]{dh12} with the CHARA array.}
\label{fig1}
\end{center}
\end{figure}
As a first validation of the asteroseismic radii and angular diameters determined, we considered a sample of 22 \textit{Kepler} targets in the main-sequence and subgiant phase. These were selected because they have solid detections of solar-like oscillations, as well as parallaxes measurements accurate to better than 10\%. In the left panel of Fig.~\ref{fig1} we show the comparison of our distance determinations with those from parallaxes. There is an excellent level of agreement between both determinations, with a weighted mean difference in the results below the 3\% level. Details on the input asteroseismic and photometric data of the stars can be found in \cite[Silva Aguirre et al. (2012)]{vsa12}.

Another test of our method comes from direct comparison of our angular diameters with those measured directly by interferometry. We applied our technique to the sample of interferometric targets studied by \cite[Huber et al.(2012)]{dh12}, finding a residual mean of 2\% between their results and our determinations (right panel in Fig.~\ref{fig1}). Interestingly enough, the \cite[Huber et al.(2012)]{dh12} sample included four red giants that confirm our method to be valid also for evolved stars. In summary, the agreement with independent measurements of parallaxes and interferometric angular diameters gives us confidence that asteroseismic determinations of stellar parameters can be accurate in radius and therefore distance to a level better than 5\%.

Knowledge of stellar properties in composite populations can be used to trace gradients across the Milky Way and help constraining the variety of processes that play a role in shaping up the Galaxy. Since the amplitudes of stellar pulsations scale proportionately with intrinsic luminosity (e.g., \cite[Kjeldsen \& Bedding (1995)]{kb95}), oscillations detected in red giants by the space missions (e.g., \cite[de Ridder et al. (2009)]{jd09}, \cite[Bedding et al. (2011)]{tb11}) offer a unique sample with a wide coverage in mass and radius to test these theories.

Using red giant asteroseismic data from the CoRoT mission, \cite[Miglio et al. (2013a), (2013b)]{am13a,am13b} carried out a first study of $\sim$2000 stars aiming at determining their stellar parameters and map the Galactic disk. The observed targets are contained in two pencil-beam structures pointing at opposite galactic longitudes, where the so-called center long run (Lrc01) reaches lower galactic latitudes than the anti-center run (LRa01, see Fig.~1 in \cite[Miglio et al. (2013a)]{am13a}).

\begin{figure}
\centering
\begin{minipage}{.5\textwidth}
  \centering
  \includegraphics[width=\linewidth]{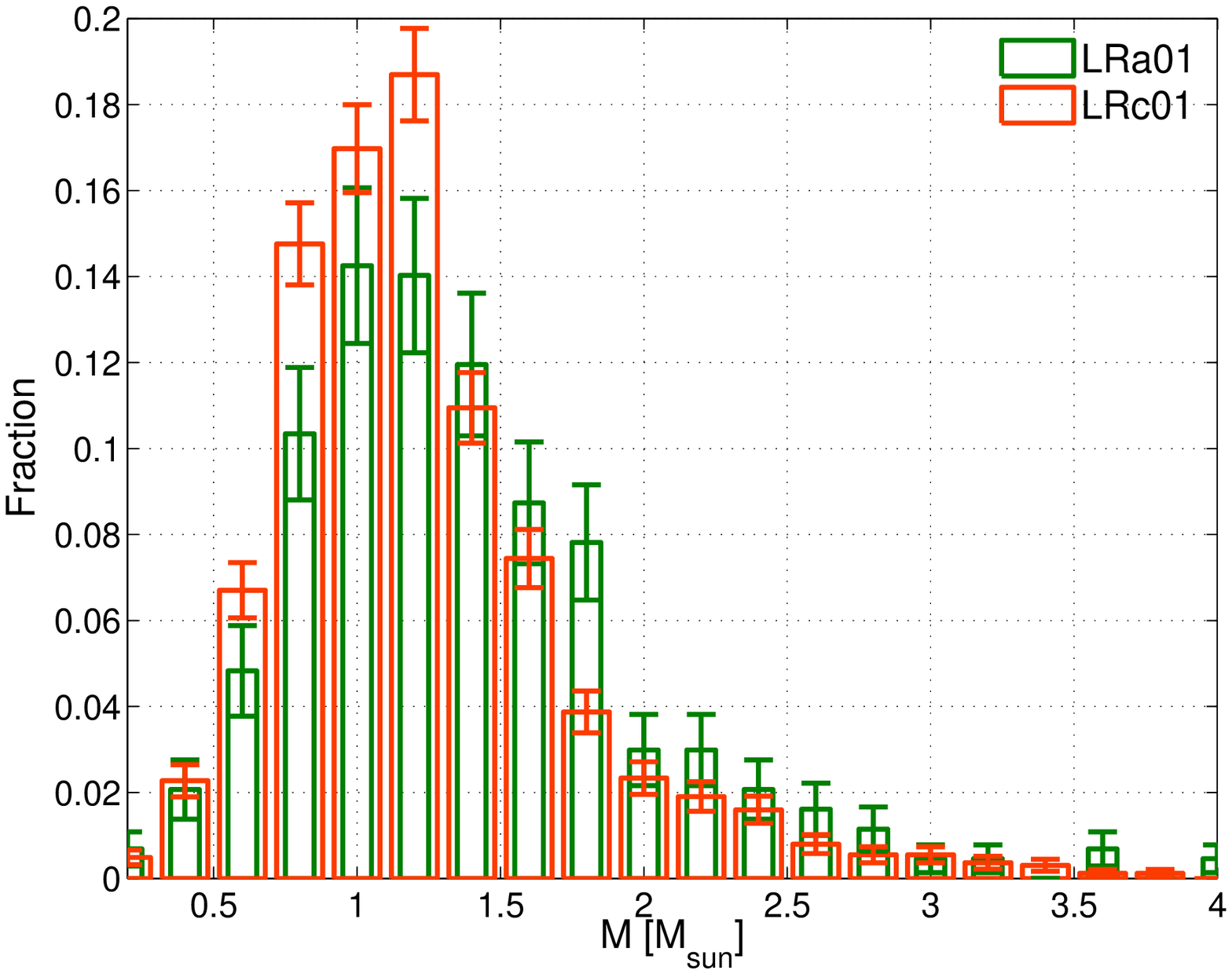}
\end{minipage}%
\begin{minipage}{.5\textwidth}
  \centering
  \includegraphics[width=\linewidth]{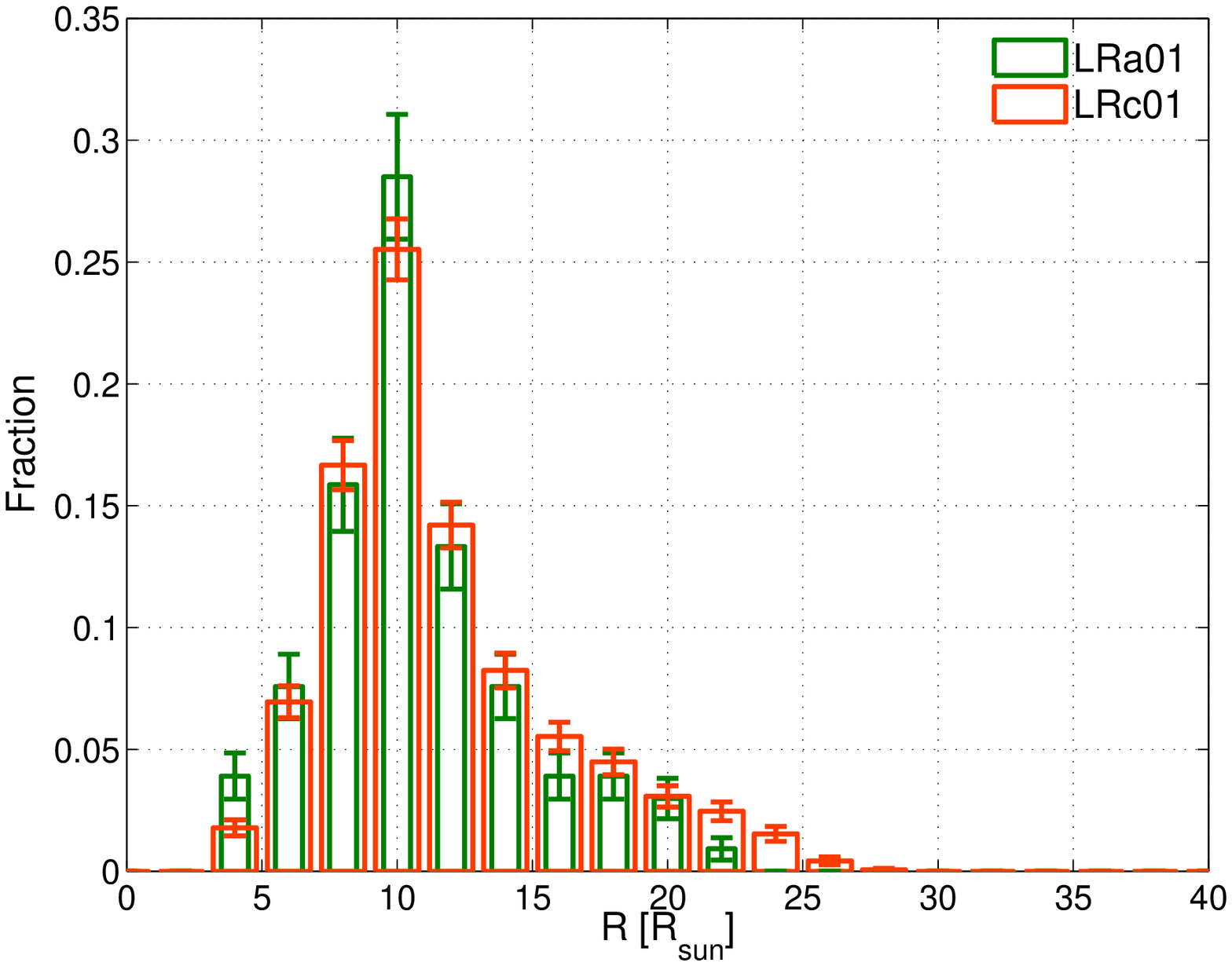}
\end{minipage}
\label{fig2}
\caption{Histograms of mass (left) and radius (right) distributions for the CoRoT red giants in the center (LRc01, red) and anti-center (LRa01, green) directions. Sample of stars taken from \cite[Miglio et al. (2013a)]{am13a}. See test for details.}
\end{figure}

Determination of masses and radii using Eqs.~\ref{eqn:mass}~and~\ref{eqn:rad}, coupled with photometric effective temperatures and bolometric corrections yielded a set of stellar parameters that help characterizing both stellar populations. In Fig.~\ref{fig2} the histograms of mass and radius distributions are shown, revealing a slight excess of low-mass stars in the center direction compared to the anti-center results. Since the center observations encompassed stars at lower galactic latitudes, the authors interpret this result as a vertical gradient in the mass distribution (and thus age) in the Galactic disk.

A natural step to follow in this line of research is to determine accurate ages for these stars. Unfortunately in the case of red giants, using only the scaling relations provides ages with uncertainties of the order of 30-40\% (\cite[Miglio et al. (2013b)]{am13b}). Better constraints can be obtained when the evolutionary stage is known to be either ascending red giant branch or core-helium burning giant directly from asteroseismic data (\cite[Bedding et al. (2011)]{tb11}). Unfortunately this information is not always readily available for all giants with detected oscillations, and other means of calibrating the scaling relations for ages of evolved stars must be employed.

In this respect, and considering that ages of red giants are mostly determined by their main-sequence lifetime, accurate determination of asteroseismic ages for unevolved stars can provide a sound base for calibrating the red giant phase. In the case of main-sequence pulsators, techniques relying on individual frequencies of oscillations have been widely studied and can set stricter constraints on ages. Since 1-D stellar models poorly reproduce the outer layers of stars, it is useful to consider combinations of frequencies that eliminate the contribution of these regions. In particular, the frequency ratios (e.g., \cite[Roxburgh \& Vorontsov (2003)]{rv03}, \cite[Silva Aguirre et al.(2011b)]{vsa11b}):
\begin{eqnarray}\label{eq:rat}
r_{01}(n)=\frac{d_{01}(n)}{\Delta\nu_{1}(n)},& & r_{10} = \frac{d_{10}(n)}{\Delta\nu_{0}(n+1)}\,,
\end{eqnarray}
where $d_{01}(n)$ and $d_{10}(n)$ are the smooth 5-points small frequency separations:
\begin{equation}\label{eq:d01}
d_{01}(n)=\frac{1}{8}(\nu_{n-1,0}-4\nu_{n-1,1}+6\nu_{n,0}-4\nu_{n,1}+\nu_{n+1,0})
\end{equation}
\begin{equation}\label{eq:d10}
d_{10}(n)=-\frac{1}{8}(\nu_{n-1,1}-4\nu_{n,0}+6\nu_{n,1}-4\nu_{n+1,0}+\nu_{n+1,1})\,.
\end{equation}

\begin{figure}[th]
\begin{center}
 \includegraphics[width=\textwidth]{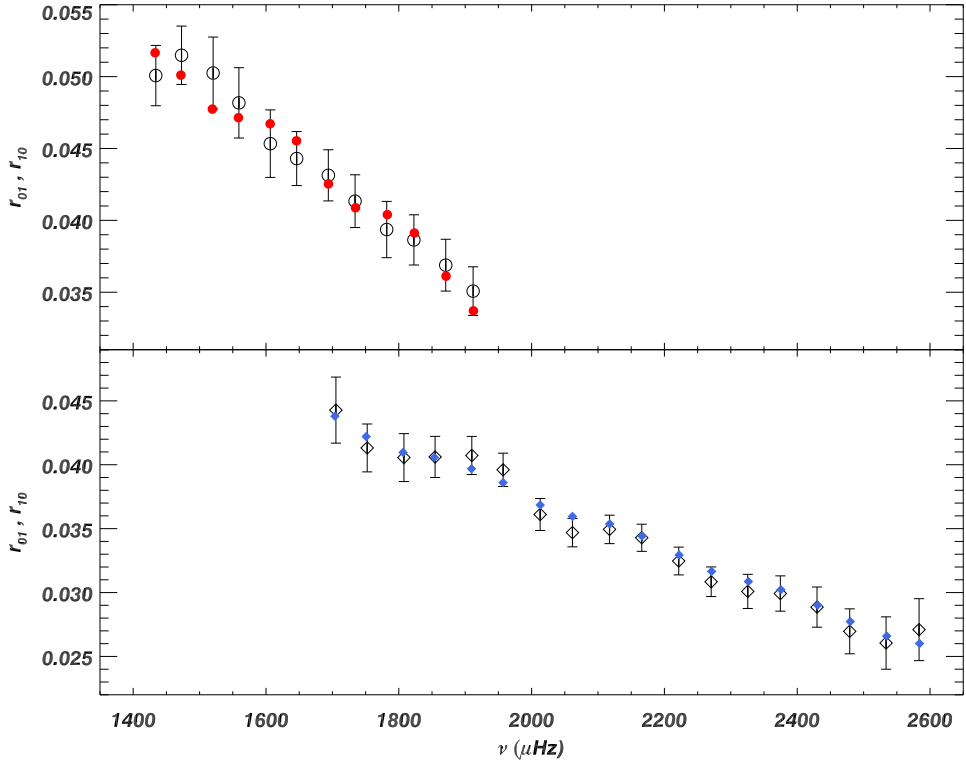} 
 \caption{Frequency ratios of the two targets studied in \cite[Silva Aguirre et al. (2013)]{vsa13}. Open symbols show the data, and filled symbols depict one of the best fitting model as an example of the quality of the fit. See text for details.}
\label{fig3}
\end{center}
\end{figure}

Taking these asteroseismic quantities into account, \cite[Silva Aguirre et al. (2013)]{vsa13} performed a detailed study of two \textit{Kepler} targets with accurate individual oscillations frequencies available. Using four different evolutionary codes coupled to three oscillation codes, the authors modelled the two targets and determined the stellar parameters by fitting the ratios $r_{01}$ and $r_{10}$ as depicted in Fig.~\ref{fig3}. Their results show that radii, masses, and ages of main-sequence field stars can be determined by this technique with a precision of $\sim$1.5\%, $\sim$4\%, and $\sim$10\%, respectively.
\section{Closing remarks}
We have shown that asteroseismic scaling relations, coupled to photometric determinations of stellar temperature and metallicity, can yield sets of accurate stellar parameters for field stars. A particularly interesting product of this technique is the determination of distances to stars, that allow us to dissect the galactic disk and study its different components. Using red giants in  different fields of the galaxy, asteroseismic analysis reveals a mass gradient as a function of galactic latitude in the Milky Way disk. This is interpreted as evidence of an age gradient that provides another constrain for models of galactic chemical evolution.

Although ages for red giants remain to be determined once information on the evolutionary state is available from seismology, calibrations of the techniques can be achieved using main-sequence oscillators as a benchmark. In this context, we have shown that detailed analysis using frequency combinations can result in ages of field stars precise to a level of $\sim$10\%. The asteroseismic revolution pioneered by the CoRoT and \textit{Kepler} missions will provide new test beds that will challenge our understanding of the Galaxy and help us reconstruct its formation history.
\acknowledgements
Funding for the Stellar Astrophysics Centre is provided by The Danish National Research Foundation (Grant agreement No. DNRF106). This research is supported by the ASTERISK project (ASTERoseismic Investigations with SONG and {\it Kepler}) funded by the European Research Council (Grant agreement No. 267864). 
%


\end{document}